# SS 433 – On flares affecting the slaved accretion disk

M. G. Bowler
Department of Physics, University of Oxford, Oxford OX1 3RH, UK.
e-mail michael.bowler@physics.ox.ac.uk

**Abstract.**
Variations in the polar angle of the precessing jets in SS 433 are negatively correlated with variations in the speed of the jets. This was established some 20 years ago, based on the analysis of archival data accumulated from 40 years ago. This curious correlation has never been explained. Here I consider a promising mechanism involving the effect of flares on the slavery of the accretion disk.

## 1. Introduction.

Since the early 1980s it has become part of the folklore concerning the unique binary SS 433 that the speed of ejection of the opposing jets (for which it is famous) is remarkably constant at 0.26c. Remarkable it may be, but not constant; the speed fluctuates about this mean by about 10%, between about 0.235 and 0.285; it has even reached as high as 0.32. These results were established first in analysis of a deep VLA image [1] and later [2] from VLBA images. Essentially the same results were obtained from archival optical spectra [3] and a succession of images taken over 75 days [4]. There are no obvious indicators of the mechanism whereby the two opposing jets are generated and here I examine more subtle features of the data in the hope of understanding a curious correlation.

## 2. Jet speeds and accretion.

There is no evidence that the two jets have speeds different from each other at any one time and indeed the Doppler shifts to the red ($z^+$) and blue ($z^-$) are remarkably (anti-) symmetric, even down to the details of nutation (or nodding) with a period of 6.28 days [4], [5] (reproduced in [6]). The implication is that the nutating and precessing central engine acts as a whole; variations in one jet reflected in the other [1].

The highest speeds are associated with flares observed in the optical spectra [4,7] and in radio images [2]. In [2] there is in addition evidence that individual bolides in the jets are brighter when travelling at the highest speeds. It now seems clear that these flares are the result of eruptions of the companion donor star, probably filling the Roche lobes, ejecting material from the binary system [8,9]. Thus observations suggest that eruptions feed more material to the accretion disk, thereby increasing the speed of ejecta in the jets.

In the optical spectra of [3] and the more recent data in [4] there is variation of jet speeds with the orbital period of 13.08 days. This also suggests that variation of the rate of transfer of material to the accretion disk affects the jet speeds; transfer maximised at periastron for an elliptical orbit. The binary orbit is elliptical (ellipticity 0.05 [10]) and the phase of the maximum of the 13.08 day component of jet speed advanced by ~ 90° in 25 years [4], suggestive of apsidal precession.

## 3. Jet speeds and the polar angle.

Remarkably, the polar angle of the jet axis relative to the precession axis is anti-correlated with the jet speed; that is, faster jets make a smaller polar angle. This was first noticed by K M Blundell in a re-analysis of the optical archival data [3] and provided a partial solution to an ancient problem. A plot of the residuals of the redshifts from the approaching jet versus those from that receding would, in the absence of velocity jitter, have a slope of -1; that is on average a distribution falling from upper left to lower right. From the archival data of [11] that slope is in fact - 0.79 and this curious observation resisted understanding until the analysis in [3]; I am not aware that much attention has been paid to this problem. That distribution is accounted for in detail in [3] with the addition of velocity jitter (anti-)correlated with the polar angle. The physical origin of this correlation must lie in some mechanism that increases the jet speeds and simultaneously reduces the polar angle. It is intuitively appealing to suppose that acceleration in the core of the disk has rotational symmetry, broken here and there, from time to time, through transverse kicks; let those impulses be independent of speed of ejection of bolides. This would generate the anti-correlation, but is contrived and *ad hoc*. Angular fluctuations in the two jets would be uncoupled; if any such effect exists, it is very small [3]. Searching for an explanation free of *ad hoc* hypotheses and consistent with complete symmetry, I find that this anti-correlation may have its origin in the effect of flaring on the slaved accretion disk in SS 433.

### 4. Slaved accretion disks.
SS 433 is a binary system in which one member is a compact object (a black hole) and the other a comparatively normal star, which feeds the accretion disk of the black hole. In such systems, if that normal star has a significant quadrupole moment and its axis of rotation is not aligned with the axis of the orbit, then the axis of rotation precesses. If this precession is copied by the disk, the disk is said to be slaved to the rotation of the normal star. In the model of [12,13] (originally devised for Her X-1) the normal star fills its Roche lobe and spills material through the $L_1$ point, preferentially when the compact object crosses the bulging equator of the normal star. The equatorial material has a component of momentum perpendicular to the orbit. If this is preserved in the fall, the disk created is tilted with respect to the orbital plane; for relatively slow precession the disk axis follows the precessing axis of the normal companion. This transfer from the $L_1$ point to the environs of the compact object is crucial to the slaved disk model; the angular momentum transferred determines the polar angle of the normal to the disk.

One interesting application to SS 433 is [14], which successfully reproduces photometric data (and concludes that the ratio of the mass of the compact object to the companion is ≥0.8); a comparatively recent observation is in [9].

### 5. Source of the correlation.
Periods of flaring - which can last 10s of days - are associated with periods of high jet velocity, with the implication that more material is then being processed by the central engine. There are other indicators; see Fig.8 of [7], for example the increase in the speed of the wind, blown from the disk. Such eruptions of the normal star that are manifested as flares will disturb and augment the stream of equatorial material. This seems likely to destroy, or at least dilute, the angular momentum component associated with motion normal to the plane of the orbit, the component responsible for slaving the accretion disk. Thus freed, the normal to the disk plane evolves closer to the normal to the orbital plane. Hence the correlation.

### 6. Discussion.
The conjectures above are qualitative and there are certain unresolved issues. First, in the slaved disk model for precession, transfer of stellar material via $L_1$ should occur preferentially every half period, at least in the absence of flaring. This determines the orientation of the accretion disk but there is no guarantee that the jet speed should exhibit a 6.5 day modulation. There is no evidence for such periodicity in the speed of the jets in [3] or in the data of [4], but prior to a flare, a periodicity of 13.08 days is clearly observed. The ellipticity of the orbit is only 0.05 [10]; it seems questionable that this alone is sufficient to impose the orbital periodicity on the jet speeds, yet it is there, together with evidence for apsidal precession. However, the amplitude of this periodic residual is only 2.5% of the average jet speed [3,4]. Finally, the correlation coefficient between excursions in jet speed and excursions in polar angle is -0.62 [4]. It is not established that some version of the scenario sketched above could account quantitatively for such a value.

**7. Conclusion.**
The effects of flaring on the slaved accretion disk proffer a mechanism that would naturally diminish the polar angle of the normal to the disk during times of unusually high jet speeds, a mechanism with the potential to explain the anti-correlation of the polar angle of the jet axis and the speed of the jets. This anti-correlation might even be further evidence that precession of the disk is slaved.